\documentclass[useAMS,usenatbib]{mn2e}
\usepackage{graphicx}
\usepackage{amsmath}
\usepackage{amssymb,latexsym,amsopn}
\title[Gravitational-wave detection and SMBH astrophysics]{Prospects for gravitational-wave detection and supermassive black hole astrophysics with pulsar timing arrays}

\author[Ravi et al.]{V. Ravi$^{1, 2}$\thanks{E-mail: v.vikram.ravi@gmail.com}, J. S. B. Wyithe$^{1}$, R. M. Shannon$^{2}$ and G. Hobbs$^{2}$
\\
$^{1}$School of Physics, University of Melbourne, Parkville VIC 3010, Australia\\
$^{2}$CSIRO Astronomy and Space Science, Australia Telescope National Facility, P.O. Box 76, Epping, NSW 1710, Australia\\
}

\begin{document}

\date{}

\pagerange{1--10} \pubyear{2014}

\maketitle

\label{firstpage}

\begin{abstract}

Large-area sky surveys show that massive galaxies undergo at least one major merger 
in a Hubble time. Ongoing pulsar timing array (PTA) experiments are aimed at measuring the gravitational wave (GW) 
emission from binary supermassive black holes (SMBHs) at the centres of galaxy merger remnants. 
In this paper, using the latest observational estimates for a range of galaxy properties and scaling 
relations, we predict the amplitude of the GW background generated by the binary SMBH population. We also predict the numbers 
of individual binary SMBH GW sources. We predict the characteristic strain amplitude 
of the GW background to lie in the range  
$5.1\times10^{-16}<A_{\rm yr}<2.4\times10^{-15}$ at a frequency of $(1\,{\rm yr})^{-1}$, with 95\% confidence. Higher values within this 
range, which correspond to the more commonly preferred choice of galaxy merger timescale, will fall within the expected sensitivity ranges 
of existing PTA projects in the next few years. In contrast, we find that a PTA consisting of at least 100 pulsars observed with next-generation 
radio telescopes will be required to detect continuous-wave GWs from binary SMBHs. We further suggest that GW memory bursts from coalescing SMBH pairs 
are not viable sources for PTAs.
Both the GW background and individual GW source counts are dominated by binaries formed in mergers between early-type galaxies of masses 
$\gtrsim5\times10^{10}M_{\odot}$ at redshifts $\lesssim 1.5$. 
Uncertainties in the galaxy merger timescale and the SMBH mass $-$ galaxy bulge mass relation dominate the uncertainty in our predictions. 

\end{abstract}


\section{Introduction}

Astrophysical gravitational waves (GWs) affect long-term timing measurements of radio pulsars \citep{ew75,s78,d79}. 
GW-induced metric perturbations at the Earth cause variations in pulse arrival times that differ between pulsars only by geometric factors. 
Hence, a specific GW signal may be directly detected in contemporaneous timing measurements of multiple pulsars. Three such 
`pulsar timing arrays' \citep[PTAs; e.g.,][]{fb90} are currently in operation: the European Pulsar Timing Array \citep[EPTA;][]{kc13}, 
the North American Nanohertz Observatory for Gravitational Waves \citep[NANOGrav;][]{mc13}, and the Parkes Pulsar Timing Array 
\citep[PPTA;][]{mhb+13}. These groups also share data as part of the International Pulsar Timing Array \citep[IPTA;][]{haa+10} consortium. 
Currently, at least 50 millisecond pulsars are observed every $2-4$ weeks, and data sets 
stretching for $5-30$\,yr exist for 34 of these pulsars \citep{m13}. Together, these timescales imply that PTAs are sensitive to GWs in the frequency band 
$10^{-9}-10^{-7}$\,Hz, which is complementary to other GW detection experiments.

The best-studied sources of GWs within the PTA frequency band are binary supermassive black holes (SMBHs). Stellar- or 
gas-dynamical evidence exists for SMBHs at the centres of 87 nearby galaxies at the time of writing \citep{kh13}, with masses $M_{\bullet}$ ranging 
between $10^{6}-10^{11}M_{\odot}$. Phenomenological models of the buildup of the cosmological mass density in SMBHs during luminous quasar phases 
for redshifts $z<5$ \citep[e.g.,][]{yt02,swm13} suggest short quasar lifetimes. This, considered together with local correlations between $M_{\bullet}$ 
and, for example, galaxy bulge mass $M_{\rm bul}$ \citep[e.g.,][]{kh13}, suggests that all massive galaxies ($M_{*}>10^{10}M_{\odot}$) 
which formed since the $z\sim2$ peak of quasar activity host SMBHs \citep[see also][]{mgg+14}. 

In the context of hierarchical structure formation, mergers are integral to the formation 
histories of massive galaxies, and evidence for interacting galaxies is seen across most of cosmic time \citep{bh92}.  Multiple SMBHs are 
expected to be found in galaxy merger products. Indeed, pairs of active 
galactic nuclei (AGN) are observed in galaxies in the late stages of mergers \citep{mm05}, with projected separations as small as 7\,pc 
\citep{rtz+06}. The central SMBHs in a pair of merging galaxies are likely to sink in the merger remnant potential well 
through dynamical friction and form a bound binary \citep[e.g.,][]{bbr80,kpb+12}. Dynamical friction becomes inefficient once stars within 
the binary orbit are ejected, and the slingshot scattering of stars on radial orbits \citep{fr76,q96,y02} or friction against circumbinary gas 
\citep{elc+04,dch+07,rds+11} is required to drive further orbital decay. At a binary component separation of $\ll1$\,pc, energy and 
angular momentum losses to gravitational wave (GW) emission can lead to SMBH-SMBH coalescence within a Hubble time 
\citep[e.g.,][]{pm63}. A few candidate binary SMBHs with such separations have been identified \citep[e.g.,][]{vln+08,bl09,ebh+12}.

The existence of a large cosmological population of binary SMBHs is thus inferred, some of which emit 
GWs in the PTA frequency band. If the orbits of all binary SMBHs radiating GWs in the PTA frequency band are circular and 
evolving under GW emission alone, the summed GW signals from all binaries together have the expected spectral form \citep[e.g.,][]{p01}:
\begin{equation}
\label{eq:c7e1}
h_{c}(f)=A_{\rm yr}\left(\frac{f}{f_{\rm yr}}\right)^{-2/3}.
\end{equation}
Here, $h_{c}(f)$ is the GW characteristic strain per logarithmic frequency unit, $f$ is the GW frequency at the Earth, 
$f_{yr}=({\rm 1\,yr})^{-1}$ and $A_{\rm yr}$ is the characteristic spectral amplitude at $f_{\rm yr}$. The summed signals from binary SMBHs 
may be collectively modelled as a GW background (GWB). GWs from individual binaries may be detectable as continuous-wave 
(CW) sources \citep[e.g.,][]{svv09,rwh+12}. Coalescing pairs of SMBHs also emit GW `memory' bursts \citep{bt87,f09}, which are abrupt, 
propagating metric changes that also affect pulsar timing measurements \citep[e.g.,][]{mcc14}.

Of these three types of GW signal, searches for a GWB provide the best constraints on models for the binary SMBH population \citep{src+13} and are perhaps 
the closest to yielding a successful detection \citep{sej+13}. \citet{src+13} used the first 
PPTA data release \citep{mhb+13} to show that 
$A_{\rm yr}<2.4\times10^{-15}$ with 95\% confidence. \citet{sej+13} suggest that the NANOGrav PTA is likely to be able to detect a GWB with 
$A_{\rm yr}=10^{-15}$ by around the year 2020. Constraints on GWs from individual binary SMBHs with the current PPTA data set   
showed that binaries with component masses $M_{\bullet}>10^{9}M_{\odot}$ and with separations of less than 0.02\,pc are unlikely to exist within 
30\,Mpc of the Earth \citep{zhw+14}. No search for memory bursts with PTA data has so far been published. Wang et al. (submitted) describe an 
unsuccessful search for memory bursts using current PPTA data.

 Predictions for GW signals from binary SMBHs are based either on physical models for 
galaxy formation and evolution which predict the binary SMBH population, 
or on directly observed quantities such as the galaxy merger rate and SMBH-galaxy scaling relations. 
The former predictions typically combine dark matter halo merger rates in the cold dark matter paradigm, analytic or 
numerical estimates of galaxy merger and binary SMBH formation timescales, prescriptions for the cosmic evolution of 
the galaxy and SMBH population and the assumption of GW-driven binary orbital evolution \citep{wl03,ein+04,svc08,rwh+12,kon+13,rws+14}. 
Once tuned to reproduce observables such as local SMBH-galaxy relations, the galaxy stellar mass function (GSMF) and colour distribution, 
and the quasar luminosity function, these models result in estimates of $A_{\rm yr}$ in the range $10^{-16} - 2\times10^{-15}$. 
The exact value depends on, for example, the assumed cosmological parameters, galaxy merger timescales and the specific 
models for SMBH formation and growth. Models for the binary SMBH population which more directly incorporate observational information \citep{jb03,s13,mop14} result in 
similar values of $A_{\rm yr}$.  While various studies suggest that individual binary SMBHs may be viable 
CW sources of GWs for PTAs \citep{svv09,rwh+12} as well as viable sources of memory bursts \citep{vl10,cj12,mcc14}, 
quantitative predictions of source counts have only been calculated for CW sources using theoretical galaxy formation models \citep{svv09}. 

In this paper, we adopt an observations-based approach towards modelling the binary SMBH population, in order to predict the range of 
GW signals in the PTA frequency band. 
A similar study by \citet{s13}, hereafter S13, combined a plethora of observational estimates of the merger rate of massive galaxies, the GSMF and 
local  
SMBH-galaxy relations to derive a range of possible GWB amplitudes. However, some of the observational quantities included in 
the S13 study do not represent the best current knowledge. Furthermore, some relevant uncertainties, such as in the possible redshift evolution of the 
SMBH-galaxy relations, were not accounted for by S13, and no predictions for individual GW sources were made. 
Besides addressing these issues, this paper builds on previous studies in the following ways:
\begin{itemize}

\item We quantify the impacts of different observational uncertainties on the amplitude of the GWB generated by binary SMBHs. We focus in 
particular on aspects of our model for the binary SMBH population for which little observational information currently exists.

\item We highlight the redshifts, masses and types of merging galaxies which result in binary SMBHs which dominate the GWB amplitude. 

\item We provide new, observations-based predictions for the counts of individual GW sources. We further present the first 
estimates for the expected numbers of detectable individual GW sources, given different PTA configurations, that are robust with respect to 
pulsar parameter fitting. 

\end{itemize}
In \S2, we outline our model, and present our results in \S3. We state the key implications of this work for PTAs in \S4. Finally, we 
discuss these results in \S5, and summarise our conclusions in \S6. Throughout 
this work, we adopt a concordance cosmology based on results from the \textit{Planck} satellite \citep{pc+13}, including 
$H_{0}=67.8$\,km\,s$^{-1}$\,Mpc$^{-1}$, $\Omega_{\Lambda}=0.692$ and $\Omega_{M}=0.308$.

\section{An empirical model for GWs from binary SMBHs}

\subsection{The SMBH-SMBH coalescence rate}

The cosmological population of binary SMBHs emitting GWs in the PTA frequency band can be characterised using the SMBH-SMBH coalescence rate. 
The numbers of binary SMBHs in different orbits are related to the coalescence rate through a continuity equation \citep{p01,rws+14} 
that includes assumptions about the rate of binary SMBH orbital evolution. 

We assume that the SMBH-SMBH coalescence rate is equivalent to the galaxy merger rate. This is justified because massive galaxy mergers are 
typically completed within a few galaxy dynamical times, whereas the timescales for two SMBHs to form a gravitationally-bound 
binary and then coalesce through losses of energy and angular momentum to their environments and GWs are much shorter \citep[e.g.,][]{bbr80,rs12}. 
In this work, we neglect systems of more than two gravitationally-interacting SMBHs resulting from multiple galaxy mergers, because we expect these to be 
rare for the high mass ratio ($\mu_{*}>1/3)$ mergers between massive ($M_{*}>10^{10}M_{\odot}$) galaxies that we consider. 
We further assume that each galaxy contains a central SMBH with a mass related to the galaxy  bulge mass. 
We use measured quantities to determine the all-sky coalescence rate of pairs of SMBHs. 
For each quantity, we define a fiducial prescription, and also describe the possible ranges over which the prescription can vary.

Similarly to S13, we express the galaxy merger rate as 
\begin{eqnarray}
\Phi_{\rm mrg}(M_{*},\mu_{*},z) &=& \frac{d^{4} N_{\rm mrg}}{d\log(M_{*})d\log(\mu_{*})dzdt} \\
\label{eq:c7e3}
&=& \frac{1}{\Gamma}\frac{dt_{p}}{dt}\frac{d^{2}N_{\rm gal}}{d\log(M_{*})dz}\frac{dP}{d\log(\mu_{*})}\Big|_{M_{*}}
\end{eqnarray}
where $N_{\rm mrg}$ is the number of mergers between two galaxies of combined stellar mass $M_{*}(1+\mu_{*})$, $\mu_{*}$ is the ratio between the smaller and larger galaxy stellar masses and $z$ is the cosmological redshift. The merger rate, 
$\Phi_{\rm mrg}$, is defined as the number of mergers per units $M_{*}$, $\mu_{*}$, $z$ and observer time $t$. In Equation~\ref{eq:c7e3}, 
$N_{\rm gal}$ is the number of galaxies across the entire sky with a given $M_{*}$ at a given $z$. This is related to the standard 
GSMF, $\Phi_{*}(M_{*},\,z)$, as
\begin{equation}
\label{eq:c7e4}
\frac{d^{2}N_{\rm gal}}{d\log(M_{*})dz}=\Phi_{*}\frac{4\pi d^{2}V_{c}}{d\Omega dz},
\end{equation}
where $\frac{4\pi d^{2}V_{c}}{d\Omega dz}$ is the sky-integrated comoving volume shell between redshifts $z$ and $z+dz$. In Equation~\ref{eq:c7e4}, 
$\frac{dP}{d\log(\mu_{*})}\big|_{M_{*}}$ is the probability density function for a galaxy merger event with mass $M_{*}$ at redshift $z$ having a mass ratio 
$\mu_{*}$, $\Gamma(M_{*},\,z)=\left(\frac{dn_{\rm mrg}}{dt_{p}}\right)^{-1}$ is the average proper time between major mergers for a galaxy 
with a mass $M_{*}$ at redshift $z$ and $\frac{dt_{p}}{dt}=(1+z)^{-1}$. Also, $\frac{dn_{\rm mrg}}{dt_{p}}\big|_{M_{*},\,z}$ is the number of mergers, 
$n_{\rm mrg}$, per unit proper time, $t_{p}$, for a single galaxy with a mass $M_{*}$ at redshift $z$.

In order to convert galaxy stellar masses to bulge masses ($M_{\rm bul}$) we distinguish between quiescent, red-sequence early-type galaxies and 
star-forming, blue-cloud late-type galaxies. We write the total GSMF as a sum of the GSMFs of 
early- ($\Phi_{\rm *,\,early}$) and late-type ($\Phi_{\rm *,\,late}$) galaxies:
\begin{equation}
\label{eq:c7e5}
\Phi_{*}(M_{*},\,z) = \Phi_{\rm *,\,early} + \Phi_{\rm *,\,late}.
\end{equation}
We relate $M_{*}$ to $M_{\rm bul}$ for early- and late-type galaxies using a scheme described in \S2.1.3.

To convert between $M_{\rm bul}$ and the central SMBH masses ($M_{\bullet}$) we use the widely-known 
$M_{\bullet}-M_{\rm bul}$ relation \citep{kh13,sgs13}. In contrast to S13, we express this relation as
\begin{equation}
\label{eq:c7e6}
\frac{dP}{d\log M_{\bullet}}=\mathcal{N}(\alpha+\beta\log M_{\rm bul},\epsilon^{2})
\end{equation}
where $\mathcal{N}(\mu,\,\sigma^{2})$ denotes a normal probability density function with centre $\mu$ and variance $\sigma^{2}$ and  $\alpha$, 
$\beta$ and the intrinsic scatter, $\epsilon$, are observationally-determined constants. It is important to account for intrinsic scatter in the 
$M_{\bullet}-M_{\rm bul}$ relation when inferring the SMBH mass function from the bulge mass function \citep[e.g.,][]{ar02}, 
because to not do so would lead to the SMBH mass function being underestimated. 

\subsubsection{The times between galaxy mergers}

Observational estimates of $\Gamma(M_{*},\,z)$ require knowledge of the fraction of galaxies, $f_{\rm gm}$, within a mass-complete sample 
at a given redshift that are undergoing mergers, and the proper time $\tau_{m}$ during which merger events can be observationally identified 
\citep[for a review, see][]{c14}. Then, $\Gamma(M_{*},\,z)=\tau_{m}/f_{\rm gm}$. In this work, we focus 
on major mergers with stellar mass ratios $\mu_{*}\geq1/3$, because these systems are likely to dominate the GW signal 
\citep[e.g.,][S13]{shm+04}. 

We consider three recent measurements of $f_{\rm gm}$ for major mergers at different redshifts in wide-area galaxy surveys, which are 
largely complete for galaxy stellar masses $M_{*}>10^{10}M_{\odot}$. These three studies fit their data to the function  
$f_{\rm gm}=a_{\rm gm}(1+z)^{b_{\rm gm}}$, where $a_{\rm gm}$ and $b_{\rm gm}$ are free parameters.  
\begin{itemize}
\item \citet{cyb09} used structural analyses of concentration, asymmetry and clumpiness (the `CAS' 
parameters) to identify systems in the process of merging among $\sim22000$ galaxies in the COSMOS and Extended Groth Strip 
surveys with $M_{*}>10^{10}M_{\odot}$ at $z<1.2$. This technique is sensitive to major mergers in particular, with mass ratios 
$\mu_{*}\gtrsim1/3$ \citep{c03}. \citet{cyb09} found $f_{\rm gm}=(0.022\pm0.006)(1+z)^{1.6\pm0.6}$.
\item \citet{xzs+12} counted galaxy pairs with projected separations between 5\,h$^{-1}$\,kpc and 20\,h$^{-1}$\,kpc from the COSMOS 
survey to estimate $f_{\rm gm}$ for $z<1$ and $\mu_{*}>0.4$. They scaled their results to include galaxy pairs for all $\mu_{*}\geq1/3$ using the 
argument that $f_{\rm gm}$ is inversely proportional to the logarithm of minimum mass ratio of the observed galaxy pair sample. 
\citet{xzs+12} found $f_{\rm gm}=(0.013\pm0.001)(1+z)^{2.2\pm0.2}$. 
\item Both the above works may suffer from incorporating small galaxy samples at low redshifts. This issue was addressed by \citet{rdd+14} using a large sample 
of galaxy pairs from the Galaxy and Mass Assembly (GAMA) survey in the redshift interval $0.05<z<0.2$. When standardised to the same projected separation, 
galaxy mass and mass ratio windows as \citet{xzs+12}, they found a substantially higher value of $f_{\rm gm}$ at these redshifts. By combining 
their results with all recent measurements of $f_{\rm gm}$ at redshifts up to 1.2, and normalising to the same projected pair separations of 
\citet{xzs+12}, \citet{rdd+14} found $f_{\rm gm}=(0.021\pm0.001)(1+z)^{1.53\pm0.08}$.
\end{itemize}
Other works have estimated $f_{\rm gm}$ with varying levels of accuracy. S13 included results from the galaxy pair studies of 
\citet{bfe+09}, \citet{dlt+09} and \citet{lli+12}. However,  \citet{bfe+09} and \citet{dlt+09} had significantly smaller pair samples than were utilised by either 
\citet{xzs+12} or \citet{rdd+14}, and \citet{lli+12} only considered major mergers of galaxies with $M_{*}>10^{11}M_{\odot}$. 

In the absence of observational estimates of the galaxy merger timescale ($\tau_{m}$) used in calculating $\Gamma(M_{*},\,z)$ from different 
measurements of $f_{\rm gm}$, we make use of theoretical predictions. However, the range of possible predictions spans a factor of three. 
\citet{kw08} used a mock galaxy catalogue from a semi-analytic model implemented within the Millennium simulation \citep{swj+05} 
to estimate $\tau_{m}$ for galaxies with different masses at different stages of merging assuming circular galaxy orbits and angular momentum loss 
through dynamical friction. However, a suite of hydrodynamic simulations of galaxy mergers conducted by \citet{ljc+08} and \citet{ljc+10}, hereafter collectively 
L08, resulted in significantly shorter merger timescales. While some authors \citep[e.g.,][]{bfe+09,rdd+14} use the estimates of \citet{kw08} to calculate 
$\Gamma(M_{*},\,z)$, others \citep[e.g.,][]{cyb09,xzs+12,c14} argue that these estimates are incorrect, at least for major mergers \citep[see footnote 15 of][]{hcb+10}. 
Analyses of cosmological hydrodynamical 
simulations combining dark matter and baryonic components suggest that the merger timescales assumed in semi-analytic models of 
galaxy formation and used by \citet{kw08} are overestimated for major mergers \citep{jjf+08}. Furthermore, 
L08 presented estimates of $\tau_{m}$ specifically calibrated to the CAS technique of \citet{cyb09} using mock galaxy images. Here, 
as a fiducial case, we only use the estimates of $\tau_{m}$ from L08, specific to estimates of $f_{\rm gm}$ from 
both galaxy pair counts and CAS analyses. These merger timescales were averaged over both field and cluster environments, and hence 
account for environmental dependencies. However, the simulation suite of L08 was not large enough to reveal significant mass- or 
redshift-dependence of $\tau_{m}$. Hence, we consider it possible that the weak dependencies on these quantities identified by \citet{kw08} may be present.

We therefore have the three following estimates of the times between galaxy mergers:
\begin{eqnarray}
\label{eq:c7e7}
\Gamma(M_{*},\,z) &=& (13.8\pm3.1)(1+z)^{-1.6\pm0.6}\,{\rm Gyr} \\
\label{eq:c7e8}
\Gamma(M_{*},\,z) &=& (19.2\pm1.5)(1+z)^{-2.2\pm0.2}\,{\rm Gyr} \\
\label{eq:c7e9}
\Gamma(M_{*},\,z) &=& (14.3\pm0.6)(1+z)^{-1.6\pm0.6}\,{\rm Gyr}, 
\end{eqnarray}
based on the work of \citet{cyb09}, \citet{xzs+12} and \citet{rdd+14} respectively. While we consider each of Equations \ref{eq:c7e7}$-$\ref{eq:c7e9} 
to be equally possible, we choose Equation~\ref{eq:c7e7} \citep{cyb09} as a fiducial prescription.  
The possible mass- and redshift-dependence of $\Gamma(M_{*},\,z)$ is given by the factor $(M_{*}/10^{10.7}M_{\odot})^{-0.3}(1+z/8)$ \citep{kw08}; 
we further consider it equally likely that this factor is present or absent, while choosing its absence as fiducial. 
Together, there are then six different possibilities for $\Gamma(M_{*},\,z)$ that we consider, each with observational uncertainties.  
We also demonstrate the effects on the GW signal from binary SMBHs of using systematically 
larger values of $\tau_{m}$ that are consistent with \citet{kw08}.

The fitting formulae in Equations \ref{eq:c7e7}$-$\ref{eq:c7e9} are consistent with results at higher redshifts \citep{c14}. 
We hence adopt these equations for $z<3$, and also assume $\frac{dP}{d\log(\mu_{*})}={\rm constant}$ \citep{xzs+12}. 
Uncertainties in $\Gamma(M_{*},\,z)$ for $z\gtrsim1$ do not significantly affect our predictions for GW signals from binary SMBHs, because, as we demonstrate, 
it appears that these signals are dominated by contributions from binary SMBHs at lower redshifts.

\subsubsection{The GSMF}

We use the latest measurements of the GSMF for $z<3$ in the range 
$10^{10}M_{\odot}\leq M_{*}\leq 10^{12}M_{\odot}$ based on the COSMOS/UltraVISTA catalogue \citep{mms+13}. \citet{mms+13} present 
GSMFs for quiescent (early-type) and star-forming (late-type) galaxies, which were identified using a colour cut. 
Utilising UV to mid-IR galaxy photometry, with 
improved sensitivity and sky-coverage over previous compilations, these authors provide the most accurate determinations of the 
early- and late-type GSMFs currently available. 

However, we still need to account for a selection of systematic errors. \citet{mms+13} use redshift, luminosity and mass measurements obtained 
through spectral energy distribution analyses. 
Assuming galaxy magnitude measurements of sufficient accuracy, systematic errors in the photometric redshifts and galaxy stellar mass measurements are dominated by how 
the stellar populations are modelled \citep[e.g.,][]{bsh+10,mlb+13,ccd+14}. Systematic errors in stellar mass measurements can lead to errors in the 
GSMF of greater than 0.6\,dex \citep{mlb+13}. \citet{mms+13} present five separate determinations of the GSMFs of early- and late-type galaxies 
using different choices for the stellar population synthesis model and star formation history, as well as expanded possibilities for galaxy metallicities and 
dust attenuation laws. We assume that each of these five GSMF determinations, for which Schechter function fits are given in Table~3 of 
\citet{mms+13}, are equally likely to be correct, but choose the default GSMF of \citet{mms+13}, given in their Table~1, as fiducial.  

The method of colour selection used to identify early- and late-type galaxies adds further systematic uncertainty to the GSMF estimates. 
For example, \citet{bsh+10} showed that edge-on dusty spiral galaxies are in fact the reddest among the galaxy population, and that more than a third of a 
red-sequence sample of galaxies could be actively star-forming objects. While \citet{bsh+10} suggest further simple morphological selections based on galaxy 
light concentrations to mitigate these effects, these data were not available in the COSMOS/UltraVISTA catalogue. Instead, we use a crude estimation 
of the uncertainty range of the GSMF caused by the colour selection from \citet{mms+13}, who presented GSMFs determined for significantly different 
colour cuts to their fiducial scheme (their Table~4). We consider this entire range of variability in the GSMF to be possible. 
We also demonstrate the effects on the resulting GW signal of possible contamination of colour-selected early-type galaxy samples with late-type galaxies 
in an extreme scenario by also performing our calculations with the early-type GSMF reduced by $1/3$.

\subsubsection{Relating $M_{*}$ to $M_{\rm bul}$}


The scheme we use to relate $M_{*}$ to $M_{\rm bul}$ for different types of galaxies is summarised as follows.
\begin{itemize}

\item Of late-type galaxies with $M_{*}>10^{10}M_{\odot}$, less than 10\% have no bulge component \citep{msp+14}; in this work, we 
assume a conservative value of 10\%. Of the others, $M_{\rm bul}/M_{*}$ is in the range $0.2\pm0.1$ \citep{lg12,msp+14,mvb14}. 

\item Early-type galaxies with $M_{*}>10^{10}M_{\odot}$ consist of a significant fraction that are best modelled with both bulges and disks, which 
are identified with the S0 (lenticular) galaxy population \citep{lg12,msp+14,mvb14}. These galaxies have values of $M_{\rm bul}/M_{*}$ which are approximately 
log-normally distributed with mean 0.7 and log-deviation $0.07$\,dex. A mild correlation between $M_{\rm bul}/M_{*}$ and $M_{*}$ may be present 
for S0 galaxies \citep{msp+14}, which we neglect in this work.   

\item For $10^{10}M_{\odot}<M_{*}\lesssim 10^{11.25}M_{\odot}$, approximately 75\% of early-type galaxies are S0s and 25\% are true ellipticals \citep{eck+11}. 
These fractions change to 55\% and 45\% respectively for larger stellar masses. 
\end{itemize}
While these results are quite approximate, and only derived for a low-redshift ($z\lesssim0.3$) galaxy sample, 
we adopt them as a fiducial scheme for relating $M_{*}$ to $M_{\rm bul}$ for $z<3$. 
This scheme is roughly consistent with that used by S13.

In the same way as accounting for scatter in the $M_{\bullet}-M_{\rm bul}$ relation raises the inferred SMBH mass function \citep[e.g.,][]{ar02}, 
the bulge mass function inferred from the GSMF will be raised given scatter in relating $M_{*}$ to $M_{\rm bul}$. 
Scatter in the $M_{\rm bul}-M_{*}$ relations can be simply combined with the scatter in the $M_{\bullet}-M_{\rm bul}$ relation by modifying Equation~\ref{eq:c7e6} as 
follows:
\begin{equation}
\label{eq:c7e10}
\frac{dP}{d\log M_{\bullet}} = \mathcal{N}\left[\log\alpha+\beta\log (M_{\rm bul}(M_{*})),\epsilon^{2}+\beta^{2}\sigma_{\rm bul}^{2}\right],
\end{equation}
where we assume $\sigma_{\rm bul}=0.1$ for both early- and late-type galaxies. In summary, the function $M_{\rm bul}(M_{*})$ in our fiducial model is defined by 
\begin{equation}
M_{\rm bul}(M_{*}) = \begin{cases}
0.2M_{*}, & {\rm for\,90\%\,of\,late\,types}, \\
0.7M_{*}, & {\rm for\,S0s} \\
M_{*}, & {\rm for\,ellipticals} 
\end{cases}
\end{equation}
We demonstrate the effects of 
possible errors in the fraction of early-type galaxies which are ellipticals by considering cases where this fraction is reduced and increased by 50\%.

\subsubsection{Relating $M_{\rm bul}$ to $M_{\bullet}$}

Despite intense interest in evincing the $M_{\bullet}-M_{\rm bul}$ relation over the last 15 years, the form of the relation remains uncertain \citep{kh13,sgs13}. 
\citet{kh13} argue that the  $M_{\bullet}-M_{\rm bul}$ relation is well modelled by a single power law for all galaxies containing classical bulges 
which include ellipticals, S0s and spirals with bulges displaying steep central light gradients. However, \citet{sgs13} find, using an extended version of 
the galaxy sample of \citet{goa+11} and independent measurements of $M_{\rm bul}$, that two power laws are required, with a break at 
$M_{\rm bul}=3\times10^{10}M_{\odot}$. 
A physical distinction between the two power laws was identified by splitting the sample into `cusp' galaxies with steep power-law central light gradients 
and galaxies where `cores', or light-deficits with respect to a cusp, are present. Cusp galaxies are typically of lower masses than core galaxies, and were 
found by \citet{sgs13} to have a steeper log-linear $M_{\bullet}-M_{\rm bul}$ relation than core galaxies. 

While we the consider the  $M_{\bullet}-M_{\rm bul}$ relations of \citet{kh13} and \citet{sgs13} equally likely, we choose the simpler relation of 
\citet{kh13} as a fiducial case. 
In Equation~\ref{eq:c7e10}, \citet{kh13} find $\alpha=-4.07\pm0.05$, $\beta=1.16\pm0.08$ and $\epsilon=0.29$. \citet{sgs13} instead find 
$\alpha=-15.37\pm0.18$ and $\beta=2.22\pm0.58$ for $M_{\rm bul}\leq3\times10^{10}M_{\odot}$ and 
$\alpha=-1.86\pm0.09$ and $\beta=0.97\pm0.14$ for $M_{\rm bul}>3\times10^{10}M_{\odot}$. As \citet{sgs13} do not estimate the intrinsic scatter, 
we assume $\epsilon=0.29$ for the entire range of $M_{\rm bul}$. 

We do not consider estimates of the $M_{\bullet}-M_{\rm bul}$ relation made substantially prior to \citet{kh13} and \citet{sgs13}. Previous estimates 
are thought to be incorrect because of systematic errors in SMBH and bulge mass estimates, the absence of recently-measured SMBH masses in brightest 
cluster galaxies, and the presence of galaxies without classical bulges in samples used to fit the relations \citep[for details, see][]{kh13}. 
Various authors infer modest redshift evolution in the $M_{\bullet}-M_{\rm bul}$ relation such that the typical ratio $M_{\bullet}/M_{\rm bul}$ may be up to a 
factor of $\sim3$ larger at $z\gtrsim2$ than the local value \citep[][and references therein]{kh13}. This can be approximately represented by letting 
$\alpha = \alpha_{0}+\log((1+z)^{K})$ with $K=1$ and $\alpha_{0}$ as above. As a fiducial case, however, we assume the conservative value of $K=0$. 

\subsection{GW signals from binary and coalescing SMBHs}

In this paper, we assume that all binary SMBHs are in circular orbits that evolve only under losses of energy and angular 
momentum to GWs. While the effects of binary SMBH environments and non-zero orbital eccentricities 
could modify the GW characteristic strain spectrum from the form in Equation~1 at frequencies up to $10^{-8}$\,Hz at the Earth, these effects 
are highly uncertain \citep[][and references therein]{rws+14}. For frequencies $f>10^{-8}$\,Hz within the PTA band (e.g., at $f=f_{\rm yr}$), the characteristic strain spectrum does indeed take the form of Equation~(1), because 
the orbits of all binaries radiating GWs at these frequencies are likely to have circularised because of GW-driven evolution. Our 
assumption allows for direct comparison with the majority of studies on this topic \citep{jb03,wl03,ein+04,svc08,rwh+12,kon+13,s13}, and for the 
GWB spectrum to be characterised by a single amplitude ($A_{\rm yr}$). 

A circular binary SMBH radiates monochromatic GWs at twice its orbital frequency. We use standard expressions from the literature for the 
rms GW strain amplitude, $h_{s}$ \citep[e.g., Equation~7 of][]{svc08} radiated by a circular binary, and the rms GW-induced sinusoidal variations to the 
pulse times of arrival (ToAs) from radio pulsars, $\sigma_{R}$ \citep[e.g., Equation~20 of][]{svv09}. Both $h_{s}$ and $\sigma_{R}$ are 
averaged over all binary orientation parameters. Following \citet{cj12}, we approximate the strain amplitude of 
a memory burst from a coalescing binary SMBH as
\begin{equation}
\label{eq:c7e14}
h_{\rm mem}=3.3\times10^{-16}\left(\frac{\eta_{\bullet}}{10^{8}M_{\odot}}\right)\left(\frac{1\,{\rm Gpc}}{D(z)}\right),
\end{equation}
where $\eta_{\bullet}=(M_{\bullet,\,1}M_{\bullet,\,2})/(M_{\bullet,\,1}+M_{\bullet,\,2})$ is the reduced mass of the coalescing binary system.

To calculate the GWB amplitude for a population of binary 
SMBHs, consider a multivariate density function, $f_{\mathbf{X}}$, for the observed binary SMBH coalescence rate, $R$, in terms of a 
$k$-component parameter vector $\mathbf{X}$ with components $X_{i}$ indexed by an integer $i$:
\begin{equation}
f_{\mathbf{X}}=\prod_{i=1}^{k}\frac{\partial[R]}{\partial X_{i}}.
\end{equation}
Following, e.g., \citet{svc08}, $A_{\rm yr}$ is given by 
\begin{equation}
\label{eq:c7e17}
A_{\rm yr} = \left[f_{\rm yr}\int...\int_{\mathbf{X}}f_{\mathbf{X}}\left(\frac{dt}{df}h_{s}^{2}\right)_{f=f_{\rm yr}}dX_{1}...dX_{k}\right]^{1/2},
\end{equation}
Here, $\frac{dt}{df}=\left(\frac{df}{dt}\right)^{-1}$ for the domains of $t$ and $f$ under consideration.

\subsection{Assembling the model}

Mergers between galaxies containing bulges with masses $M_{*}$ and $M_{*}\mu_{*}$ come in nine types, because the galaxies with each 
mass may be either elliptical, S0 or late-type. In each case, a different prescription is required to identify the bulge masses of the 
merging galaxies, and hence the masses of the SMBHs in the merging galaxies. Consider a merger between a galaxy of type $i$, 
with mass $M_{*}$, and a galaxy of type $j$, with mass $M_{*}\mu_{*}$,  where $i$ and $j$ each denote either an elliptical, S0 or  
late-type galaxy. The fraction of cases where this merger will occur is given by $\frac{\Phi_{*,\,j}}{\Phi_{*}}$, where $\Phi_{*}$ is given by 
Equation~\ref{eq:c7e5} and the mass functions are evaluated at a mass $M_{*}\mu_{*}$. For early-type galaxies, $\Phi_{*,\,j}$ is specified according 
to the fractions of ellipticals and S0s at different masses, and for late-type galaxies, $\Phi_{*,\,j}$ is simply the fraction which contain bulges. 
The SMBH masses corresponding to the galaxies of types $i$ and $j$, $M_{\bullet,\,i}$ and $M_{\bullet,\,j}$ respectively, are described by the probability 
density function in Equation~\ref{eq:c7e10} for $M_{\rm bul}$ given by $M_{{\rm bul},\,i}(M_{*})$ and $M_{{\rm bul},\,j}(M_{*}\mu_{*})$ respectively. Hence, in order to calculate 
$A_{\rm yr}$, we combine Equations \ref{eq:c7e3}, \ref{eq:c7e4}, \ref{eq:c7e5}, \ref{eq:c7e10} and \ref{eq:c7e17} as follows:
\begin{align} 
\label{eq:c7e18}
\begin{split}
\frac{A_{\rm yr}^{2}}{f_{\rm yr}} =  &  \int_{\log(10^{10}M_{\odot})}^{\log(10^{12}M_{\odot})}d\log M_{*}\int_{0}^{3}dz\int_{\log(1/3)}^{0}d\log\mu_{*} \\
& \times \int_{-\infty}^{\infty}d\log M_{\bullet,\,i}\int_{-\infty}^{\infty}d\log M_{\bullet,\,j}  \\
& \times\frac{4\pi d^{2}V_{c}}{d\Omega dz}\frac{1}{\Gamma}\frac{dP}{d\log(\mu_{*})}\frac{dt_{p}}{dt}\left(\frac{dt}{df}\right)_{f=f_{yr}} \\
& \times \frac{dP}{d\log M_{\bullet,\,i}}\Big|_{M_{*}} \frac{dP}{d\log M_{\bullet,\,j}}\Big|_{M_{*}\mu_{*}}  \\
&\times \sum_{i,\,j}\Phi_{*,\,i}\frac{\Phi_{*,\,j}}{\Phi_{*}}h_{s}^{2}(M_{\bullet,\,i},\,M_{\bullet,\,j},\,z,\,f_{\rm yr}). \end{split}
\end{align}
We evaluate this integral numerically by summing over the integrand in bins of $\log M_{*}$, $z$, $\log\mu_{*}$, $\log M_{\bullet,\,i}$, 
and $\log M_{\bullet,\,j}$. To determine the predicted numbers of CW sources, 
we count the numbers of individual binary SMBHs in each bin with different values of $h_{s}$ radiating GWs at $f_{\rm yr}$ within a nominal bandwidth 
of $\Delta f=(10\,{\rm yr})^{-1}$. We also record the rate of memory bursts 
in each bin with corresponding amplitudes $h_{\rm mem}$. These latter operations are equivalent to numerically evaluating the 
conditional densities of GW sources in terms of $h_{s}$ and $h_{\rm mem}$.

Equation~\ref{eq:c7e18} builds on the approach of S13 in two ways. First, we account for the effects of intrinsic scatter in the $M_{\bullet}-M_{\rm bul}$ 
relation and in relating $M_{*}$ to $M_{\rm bul}$. We also attempt to match the numbers of galaxy mergers of different types to the measured GSMFs, 
rather than assuming the same galaxy pair fractions for all types of mergers. 

%
%
%
%
%
%
%
%

\section{Results}


\subsection{The GWB amplitude}

We first calculated $A_{\rm yr}$ using Equation~\ref{eq:c7e18} given the fiducial prescriptions for $\Gamma(M_{*},z)$, the GSMF, the scheme relating 
$M_{*}$ and $M_{\rm bul}$ and the $M_{\bullet}-M_{\rm bul}$ relation, as detailed in sections 2.1.1$-$2.1.4. The resulting fiducial value for 
$A_{\rm yr}$ was $1.3\times10^{-15}$. We then identified the possible ranges of $A_{\rm yr}$ consistent with the observational uncertainties in 
each of $\Gamma(M_{*},z)$, the GSMF and the $M_{\bullet}-M_{\rm bul}$ relation alone. This was accomplished by generating 600 realisations 
of $A_{\rm yr}$ with the parameters of a single one of these quantities randomised and with the other terms in Equation~\ref{eq:c7e18} 
held fixed at their fiducial values. The process was then repeated with randomisation individually in the other two quantities. 

\begin{figure}
\centering
\includegraphics[angle=-90,scale=0.75]{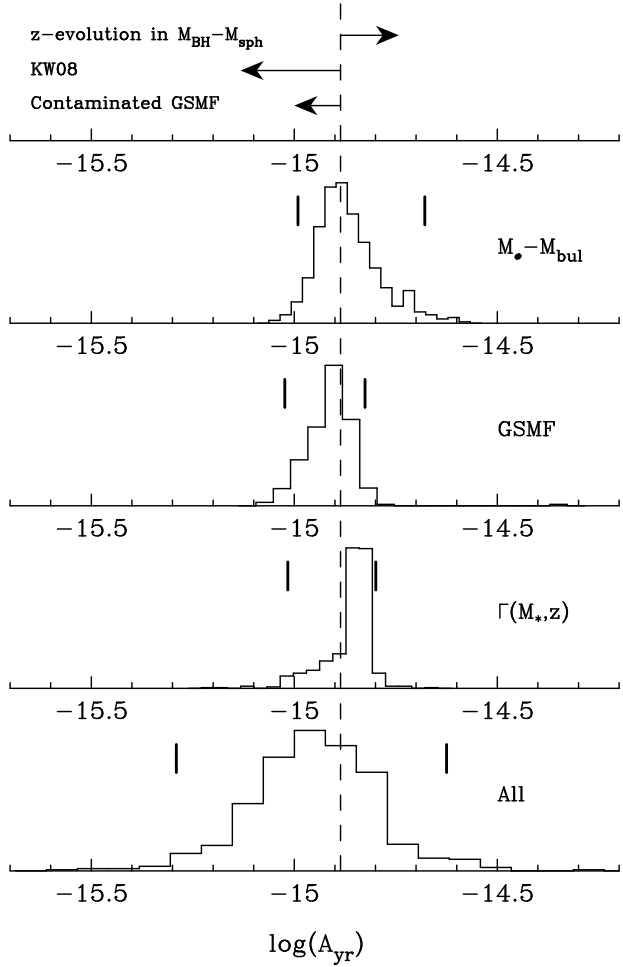}
\caption{Depiction of uncertainties in the value of $A_{\rm yr}$ calculated using Equation~\ref{eq:c7e18}. The 
standardised histograms labelled `$\Gamma(M_{*},z)$', `GSMF' and `$M_{\bullet}-M_{\rm bul}$' show the distributions of 
600 realisations of $A_{\rm yr}$ given randomisation over the prescriptions for the respective quantities alone. 
 The vertical dashed line indicates the value of 
$A_{\rm yr}=1.3\times10^{-15}$ resulting from the fiducial prescriptions for all quantities in Equation~\ref{eq:c7e18}. The three arrows at the top of 
the Figure show how much this fiducial value varies given possible systematic uncertainties in our model. From the bottom, the arrowheads indicate 
the values of $A_{\rm yr}$ corresponding to a possibly contaminated early-type GSMF, galaxy merger timescales consistent with \citet{kw08} and 
redshift-evolution in the normalisation of the $M_{\bullet}-M_{\rm bul}$ relation respectively (see text for details). The standardised 
histogram labelled `All' shows the distribution of 600 realisations of $A_{\rm yr}$ given randomisations over all uncertainties considered in this paper. 
For all four histograms, the 2.5\% and 97.5\% percentiles are shown as thick vertical bars.}
\label{fig:c7f1}
\end{figure}

Histograms of the resulting three samples of realisations of $A_{\rm yr}$ are shown in the middle three panels of 
Figure~\ref{fig:c7f1}, along with the fiducial value of $A_{\rm yr}$ (as a vertical dashed line). 
While the possible ranges of $A_{\rm yr}$ given observational uncertainties in $\Gamma(M_{*},z)$ and the GSMF are roughly equivalent, 
observational uncertainty in the $M_{\bullet}-M_{\rm bul}$ relation results in a slightly larger range of possible $A_{\rm yr}$ values.

\begin{figure}
\centering
\includegraphics[angle=-90,scale=0.54]{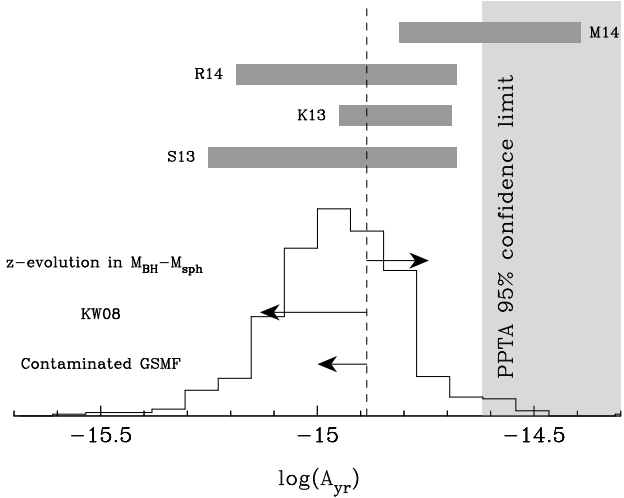}
\caption{Comparison between our predictions for $A_{\rm yr}$ and those from other works. The histogram is identical to that labelled `All' in Figure~\ref{fig:c7f1}. 
The vertical dashed line indicates our fiducial prediction of 
$A_{\rm yr}=1.3\times10^{-15}$, and the arrows indicate systematic uncertainties in $A_{\rm yr}$ (see the caption of Figure~\ref{fig:c7f1} for details). The 
dark grey horizontal bars show 68\% confidence intervals for $A_{\rm yr}$ predicted by S13, \citet{kon+13}, labelled `K13', 
\citet{rws+14}, labelled `R14', and \citet{mop14}, labelled `M14'. The light grey shaded area indicates the 95\% confidence PPTA upper limit on the 
GWB amplitude, set at $A_{\rm yr}=2.4\times10^{-15}$.}
\label{fig:c7f2}
\end{figure}

We also considered the effects of adopting four modifications to the fiducial model relating to parameters for which current observational constraints are poor. 
These modifications, which were introduced in Subsections 2.1.1$-$2.1.4, result in the following values of $A_{\rm yr}$:
\begin{enumerate}

\item When we decrease the early-type GSMF by a factor of $1/3$ to simulate an extreme case of contamination of colour-selected early-type galaxy samples 
by late-type galaxies (e.g., edge-on spirals), we obtain $A_{\rm yr}=10^{-15}$. This represents a decrease of 0.12\,dex over the fiducial model.

\item When we adopt the massive galaxy merger timescale from \citet{kw08} with the associated mass- and redshift-dependence, 
rather than from numerical simulations of galaxy mergers (L08), we obtain $A_{\rm yr}=7.4\times10^{-16}$. This represents a decrease of 
0.24\,dex over the fiducial model. 

\item When we introduce a redshift-dependent normalisation, $\alpha$, of the $M_{\bullet}-M_{\rm bul}$ relation with $K=1$ such that the normalisation 
is a factor of three greater at $z=2$, we obtain $A_{\rm yr}=1.8\times10^{-15}$. This is an increase of 0.14\,dex over the fiducial model.

\item Finally, when we explore the effects of either reducing or increasing the fraction of early-type galaxies which are ellipticals by 50\%, we obtain an 
associated variation in $A_{\rm yr}$ of 5\% (0.02\,dex).  

\end{enumerate}
The first three modifications are clearly significant, as compared to the fourth: we depict the resulting values of $A_{\rm yr}$ in the top panel of Figure~\ref{fig:c7f1}. 
While the effects of modifications ($i$) and ($iii$) are comparable in magnitude, adopting the galaxy merger timescales of \citet{kw08} makes a large difference to the prediction of $A_{\rm yr}$. This is 
expected, because the  \citet{kw08} merger timescales are roughly a factor of three longer than those of L08. Indeed, modification ($ii$) results in a 
value of $A_{\rm yr}$ that is lower than the 2.5\% percentile of the distributions of $A_{\rm yr}$ values given the three observational uncertainties considered so far. 

As an illustration of the full range of possible values of the GWB amplitude given the uncertainties considered in $\Gamma(M_{*},z)$, the GSMF 
and the $M_{\bullet}-M_{\rm bul}$ relation combined with modifications $(1)-(3)$ listed above, we generated a new sample of 600 realisations of 
$A_{\rm yr}$. In this case, we simultaneously randomised over $\Gamma(M_{*},z)$, the GSMF and the $M_{\bullet}-M_{\rm bul}$ relation as described 
above, and also (\textit{i}) decreased the early-type GSMF by a factor uniformly drawn from the interval $[0,1/3]$, (\textit{ii}) set the galaxy merger timescale at a 
value uniformly drawn between the predictions of L08 and \citet{kw08} (neglecting any mass- or redshift- dependence), and (\textit{iii}) 
set the redshift-evolution index $K$ of the normalisation of the $M_{\bullet}-M_{\rm bul}$ relation to a number uniformly drawn from the interval 
$[0,1]$. A histogram of the resulting sample of realisations of $A_{\rm yr}$ is shown in the bottom panel of Figure~\ref{fig:c7f1}, labelled `All'. 
The long tail to lower values of $A_{\rm yr}$, which is not reflected in the other histograms, is caused specifically by the inclusion of uncertainties in the 
galaxy merger timescale and in the early-type GSMF. The magnitudes of these effects on $A_{\rm yr}$ are indicated by the arrows at the top of Figure~\ref{fig:c7f1}.
The 95\% confidence interval on $A_{\rm yr}$, considering all uncertainties, is $5.1\times10^{-16}<A_{\rm yr}<2.4\times10^{-15}$.

We next compare our results for $A_{\rm yr}$ with earlier predictions. In Figure~\ref{fig:c7f2}, we again show the histogram of realisations 
of $A_{\rm yr}$ corresponding to 
randomisation over all uncertainties, as well as the values of $A_{\rm yr}$ corresponding 
to modifications ($i$) to ($iii$) listed above. Above these, we show the 68\% confidence intervals on $A_{\rm yr}$ from four recent, independent 
models for the binary SMBH population \citep[S13;][]{kon+13,rws+14,mop14}. 
The predictions that we consider all account for the most recent determinations of the $M_{\bullet}-M_{\rm bul}$ relation \citep{kh13,sgs13}. 

The range of possible values of $A_{\rm yr}$ predicted by S13 is consistent with (albeit somewhat broader than) the range we predict given all 
uncertainties that we consider in this paper. Both the present work and S13 attempt to synthesise all uncertainties 
in quantities relevant to characterising the SMBH-SMBH coalescence rate, and use the same underlying model assumptions to predict the GWB amplitude. 
Our range of predictions is less extended than that of S13 because of the greater uncertainty assumed by S13 in the GSMF and the galaxy merger rate.\footnote{Some methodological differences also exist between the present work and S13 in how different realisations of $A_{\rm yr}$ were obtained.}
 
A semi-analytic approach  \citep{gwb+11} was used by \citet{rws+14} to predict SMBH-SMBH coalescence rates within the Millennium 
simulation \citep{swj+05}, coupled with prescriptions for binary SMBH orbital evolution in stellar environments \citep{s10}. The 
results of \citet{rws+14} indicate that the characteristic strain spectrum may be attenuated relative to the case of circular binary orbits and  
GW-driven evolution at frequencies $f\lesssim10^{-8}$\,Hz. However, the 68\% confidence interval on the characteristic strain spectral 
amplitude at a frequency of $f_{\rm yr}$ is consistent with the range of values of $A_{\rm yr}$ we find. 

\begin{figure*}
\centering
\includegraphics[angle=-90,scale=0.72]{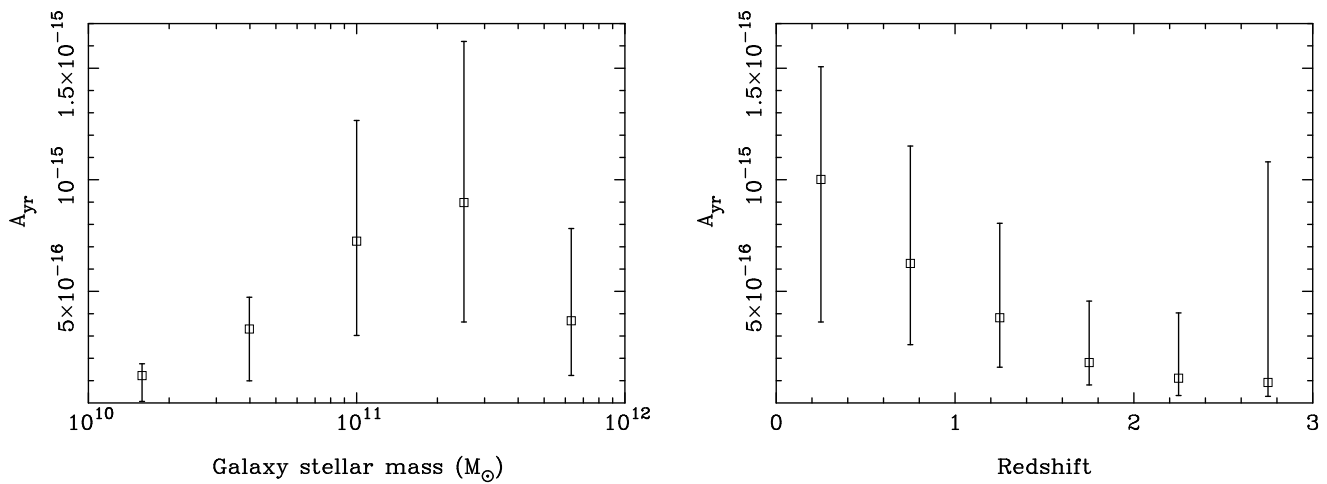}
\caption{\textit{Left: } Values of $A_{\rm yr}$ from binary SMBHs created in major mergers involving galaxies of different 
stellar masses. The squares indicate the fiducial model result, and the vertical error bars indicate 95\% confidence 
intervals from 600 realisations of the model with all uncertainties that we account for. 
\textit{Right: } Values of $A_{\rm yr}$ from binary SMBHs in six redshift bins in the interval $0<z<3$. The squares and error bars are as in 
the left panel. The redshift intervals correspond to the ranges within which the GSMF was evaluated by \citet{mms+13}.}
\label{fig:c7f3}
\end{figure*}

The prediction of 
\citet{kon+13} is derived from hydrodynamic numerical galaxy formation simulations in cluster and field environments, but may be biased 
relative to semi-analytic galaxy formation models implemented in large-volume numerical dark matter simulations because of the 
specific choice of overdense and underdense regions to study. However, the prediction of \citet{kon+13} naturally includes a particiularly sophisticated  
treatment of galaxy merger timescales. 

\citet{mop14} suggest a model for the binary SMBH population which includes the assumption that all evolution in the early-type GSMF at $z<1$ 
is driven by galaxy mergers; however, their predicted GWB amplitude appears to be inconsistent with current PTA constraints \citep{src+13}. This model would 
necessarily include a shorter galaxy merger timescale than that predicted by L08 in order to maintain consistency with the observed numbers of 
merging galaxies. Overall, besides the study of \citet{mop14}, it is encouraging that different 
models appear to agree on the amplitude of the characteristic strain spectrum from binary SMBHs. In particular, the upper ends of most predicted 
ranges of $A_{\rm yr}$ all appear to be consistent.

In Figure~\ref{fig:c7f2}, we also depict the best existing 95\% confidence upper limit on $A_{\rm yr}$ from \citet{src+13} as a shaded region. 
Some realisations of $A_{\rm yr}$ given observational uncertainties in our model are inconsistent with this upper limit. However, the upper limit is 
generally consistent with our model given all uncertainties.

In Figure~\ref{fig:c7f3}, we plot the values of $A_{\rm yr}$ predicted by the fiducial model in different ranges of $M_{*}$ (left panel) and 
$z$ (right panel). We also show the 95\% confidence intervals on these values given all uncertainties we consider. The galaxy mass ranges 
correspond to the values of $M_{*}$ of the larger galaxies in mergers. The dominant contributions to the GWB are from binary SMBHs formed 
in mergers involving galaxies with $M_{*}\gtrsim5\times10^{10}M_{\odot}$, and from binary SMBHs at redshifts $z\lesssim1.5$. 
The confidence intervals that we provide further 
suggest that contributions to the GWB from outside these ranges are not significant.\footnote{While the most massive galaxies do not appear to contribute significantly to the GWB, it is apparent from, e.g., Figure~6 of \citet{mms+13} 
\citep[see also][]{bdl+12} that the Schechter function fits to the early-type GSMFs under-predict the observed GSMF at masses $M_{*}\gtrsim10^{11.5}$.} 
Finally, binary SMBHs created in mergers involving at least one late-type galaxy correspond to $A_{\rm yr}=4.7\times10^{-16}$, whereas mergers 
involving only early-type galaxies correspond to $A_{\rm yr}=1.2\times10^{-15}$. Hence,  within our model, 
the GWB is likely to be dominated by galaxy mergers involving only early-type galaxies (S0s and ellipticals).

\subsection{Individual GW sources: continuous waves and memory bursts}

\begin{figure*}
\centering
\includegraphics[angle=-90,scale=0.72]{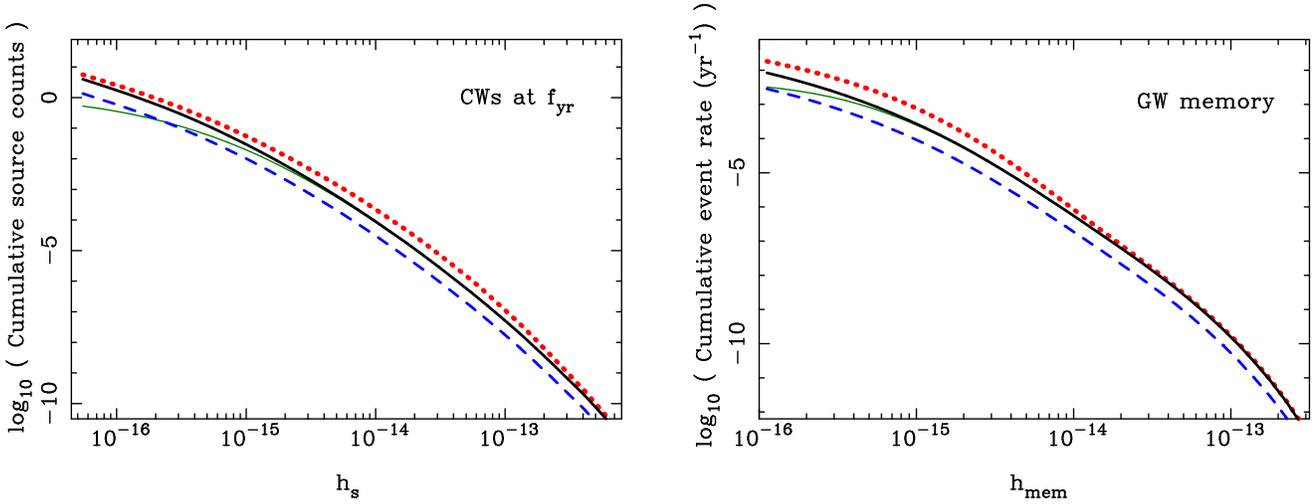}
\caption{\textit{Left:} the counts of individual sources at and above given GW strain amplitudes ($h_{s}$) 
at a GW frequency of $f_{\rm yr}$ in a frequency bin of width $\Delta f=(10\,{\rm yr})^{-1}$. \textit{Right:} the numbers of GW memory 
bursts per year at and above given strain amplitudes ($h_{\rm mem}$, Equation~\ref{eq:c7e14}). In both panels, the results of the fiducial model 
are shown as thick black solid curves, the results from a model with maximal redshift evolution in the $M_{\bullet}-M_{\rm bul}$ relation 
($K=1$ corresponding to $\alpha$ in Equations \ref{eq:c7e6} and ref{eq:c7e10} increased by a factor of three at $z=2$) are shown as dotted red curves, 
and the results from a model with galaxy merger timescales consistent with \citet{kw08} are shown as blue dashed curves. The green 
thin solid curves represent source counts for the fiducial model evaluated with the restrictions $z<1$ and $M_{*}>10^{11}M_{\odot}$. }
\label{fig:c7f4}
\end{figure*}

In the process of evaluating Equation~\ref{eq:c7e18}, we also calculated the numbers of individual binary SMBHs that produce  
monochromatic (CW) GW signals, along with the numbers of GW memory bursts emitted during SMBH-SMBH coalescence 
events. We counted individual binaries emitting GWs at frequencies $f=f_{\rm yr}$ in a frequency bin of width 
$\Delta f=(10\,{\rm yr})^{-1}$ and evaluated the numbers of binaries with different GW strain amplitudes $h_{s}$.  
These results are shown in the left panel of Figure~4 for the fiducial 
model as well as for two variations to the fiducial model (modifications (ii) and (iii) listed above). We also show results for the 
fiducial model while restricting the source counts to binaries at redshifts $z<1$ and with the more massive progenitor galaxy mass 
$M_{*}>10^{11}M_{\odot}$. The restricted source counts are identical to the full source counts for $h_{s}\gtrsim2\times10^{-15}$. 
From \citet{svv09}, the 
characteristic amplitude of the sinusoidal ToA variations induced by a binary SMBH with strain amplitude $h_{s}$ at $f=f_{\rm yr}$, over a 10\,yr 
observation, is $\sigma_{R}=21 (h_{s}/10^{-15})$\,ns. 

Scaling these CW source counts to other GW frequencies is non-trivial. The GW strain amplitude of a binary SMBH radiating at a frequency $f$ 
can be expressed as $h_{s}=h_{s,\,{\rm yr}}(f/f_{\rm yr})^{2/3}$, where $h_{s,\,{\rm yr}}$ is the strain amplitude radiated by that binary at a frequency $f_{\rm yr}$. 
Furthermore, the total number of binaries per unit frequency radiating GWs at a frequency $f$ is related to the number of binaries per unit frequency radiating 
GWs at $f_{\rm yr}$ by the factor $(f/f_{\rm yr})^{-11/3}$, assuming GW-driven binary orbital evolution. Then, the number of binaries 
per unit frequency emitting GWs at or above a strain amplitude of $h_{s}$, at a frequency $f$, may be written as $n(f,h_{s})=n(f_{\rm yr},h_{s}(f/f_{\rm yr})^{-2/3})(f/f_{\rm yr})^{-11/3}$. For example, while the fiducial model predicts $\sim$$10^{-2}$ CW sources with $h_{s}\geq 10^{-15}$ in a frequency bin of width 
$\Delta f=(10\,{\rm yr})^{-1}$ at $f=f_{\rm yr}$, this prediction changes to $\sim$0.1 sources at $f=f_{\rm yr}/5$ with 
$h_{s}\geq 10^{-15}$ in the same frequency bin width.

We can hence directly compare our predicted CW source counts with the work of \citet{svv09}. These authors considered a wide variety of SMBH growth scenarios 
within the framework of a semi-analytic model for galaxy formation \citep{bdt07} implemented in the Millennium simulation results \citep{swj+05}. We directly compare predictions for the number of binary SMBHs inducing ToA variations with characteristic amplitudes $\sigma_{R}\geq30$\,ns. 
For consistency, we consider an observation time span of $T=5$\,yr and GW frequencies $f>3\times10^{-9}$\,Hz, and integrate over the 
number of sources per unit frequency with $\sigma_{R}\geq30$\,ns in the range $3\times10^{-9}-10^{-7}$\,Hz (integrating to higher frequencies 
does not significantly alter our results). 
We neglect the issue of whether these signals are 
resolvable given the presence of a GWB. We predict 0.6 CW sources with $\sigma_{R}\geq30$\,ns for our fiducial model, 0.1 CW sources 
for a pessimistic model assuming the galaxy merger timescales of \citet{kw08}, and 1.2 CW sources 
for our optimistic model with significant redshift-evolution in the $M_{\bullet}-M_{\rm bul}$ relation. 
\citet{svv09} predict between 0.05 and 3 such sources (their Figure~3), which is consistent with our results.

We also predicted the numbers of binary SMBH coalescence events per observed year 
at or above a given GW memory burst amplitude, $h_{\rm mem}$ 
(see Equation~12) for $h_{\rm mem}>10^{-16}$. The results are shown in the right panel of Figure~4, again for the fiducial 
model and two variations to this model. We also again show results for the 
fiducial model with the restrictions of $z<1$ and $M_{*}>10^{11}M_{\odot}$; for $h_{\rm mem}\gtrsim6\times10^{-16}$, the restrictions make 
no significant difference. 

In summary, the expected numbers of individual GW sources predicted by our empirical binary SMBH model are small. At most $\sim1$ 
CW source is expected to induce ToA variations with characteristic amplitudes $\geq30$\,ns over a 5\,yr observation time span. 
Also, approximately one GW memory burst with $h_{\rm mem}>5\times10^{-16}$ is 
expected every 1000\,yr.

\section{Implications for GW detection with PTAs}

\subsection{The GWB from binary SMBHs} 

The future sensitivities of PTAs to the GWB are the subjects of ongoing research \citep[e.g.,][]{sej+13,mtg14,hdm+14}. For example, future pulsar 
observing systems and cadences, new pulsar discoveries, the effects of the interstellar medium and pulsar timing noise characteristics, 
all of which significantly affect PTA sensitivities, are difficult to forecast because of a lack of quantitative, predictive models. 
An idealised treatment of the problem by 
\citet{sej+13} suggests that, for the NANOGrav collaboration, a GWB with amplitude $A_{\rm yr}=10^{-15}$ may be 
detectable before the year 2020. We note that \citet{sej+13} assumed that the GWB characteristic strain spectrum has the power law form given 
in Equation~\ref{eq:c7e1}. We find in this paper that the GWB amplitude is likely to be in the range 
$5.1\times10^{-16}<A_{\rm yr}<2.4\times10^{-15}$ with 95\% confidence. 
If the GWB amplitude were to lie in the upper part of this range, as is expected given the more commonly preferred major galaxy merger timescale (L08), we 
suggest that detecting a GWB from binary SMBHs is indeed an attainable, short-term goal for PTAs.\footnote{If we calculate the range of 
possible $A_{\rm yr}$ values given all uncertainties, while assuming the L08 galaxy merger timescales, we find $A_{\rm yr}>9\times10^{-16}$ with 
 95\% confidence.} 

What can PTA upper limits on or detections of the GWB reveal about the determinants of the GWB amplitude?
The GWB may be parameterised by a single number, $A_{\rm yr}$ \citep[at least at GW frequencies $f\gtrsim10^{-8}$\,Hz;][]{rws+14},
the value of which is dependent on myriad quantities. Useful information can be gleaned if one of these quantities is particularly unconstrained otherwise. 
For example, if we remain agnostic with respect to the galaxy merger timescale, 
a particular value of this timescale would correspond to a range of possible GWB amplitudes given our knowledge of all the other 
determinants of $A_{\rm yr}$. Then, a PTA constraint on $A_{\rm yr}$ would correspond to a constraint on the galaxy merger timescale, 
given the assumptions inherent in our model. Through such exercises, PTAs could directly impact our understanding of galaxy 
and SMBH growth, in a more general sense than by testing specific GWB models using PTA data. We leave a demonstration of such techniques for 
future work.


\subsection{CW signals from individual binary SMBHs} 

Future PTA observations with planned telescopes such as the Five Hundred Metre Aperture Spherical Telescope 
\nocite{lnp13}(FAST, Li, Nan \& Pan 2013) and the Square Kilometre Array \citep[SKA,][]{ckl+04} may include up to 100 pulsars 
with timing noise standard deviations of $\sim$100\,ns \citep{l13,hdm+14}. 
\citet{esc12} constructed theoretical PTA sensitivity curves using simulated data sets with both 100 arbitrarily-located pulsars or 
17 pulsars at the locations of the best-timed pulsars observed by the NANOGrav collaboration, in all cases with timing noise standard deviations 
of 100\,ns and 5\,yr observation times. These sensitivity curves, shown in their Figure~5, represent the values of $h_{s}$ at different frequencies 
at which the probability of a false detection was less than $10^{-4}$ in 95\% of realisations of their simulated data sets. 
Importantly, the sensitivity curves were averaged over all source positions and orientations, and account for pulsar parameter fitting. 
We predict the numbers of detectable sources for PTAs with these sensitivity curves by evaluating the following integral:
\begin{equation}
N_{\rm detect} = \int_{({\rm 10\,yr})^{-1}}^{10^{-7}\,{\rm Hz}} \frac{dF[h_{\rm sens}(f)]}{df}df,
\end{equation}
where $h_{\rm sens}(f)$ is the sensitivity curve and $\frac{dF(h_{\rm sens})}{df}$ is the predicted number of sources with strain amplitudes 
$h_{s}\geq h_{\rm sens}(f)$ per unit frequency at a frequency $f$. The sensitivities of PTAs to CW sources are generally poor for frequencies 
$f\gtrsim 10^{-7}$\,Hz  and few sources are expected at these frequencies. 

Using our predictions for the numbers of CW sources, we evaluate $\frac{dF(h_{\rm sens})}{df}$ by scaling the predictions 
as described in \S3.2. Then, for the fiducial model and for the two sensitivity 
curves of \citet{esc12} corresponding to their coherent $\mathcal{F}$-statistic, we obtain predictions of 0.07 and 1.3 detectable sources for 
the 17- and 100-pulsar cases respectively. For the restricted fiducial model, corresponding only to sources with $z<1$ and $M_{*}>10^{11}M_{\odot}$ 
these reduce marginally to 0.06 and 1 source respectively. 
For the optimistic case with strong redshift-evolution of the $M_{\bullet}-M_{\rm bul}$ relation,
we obtain predictions of 0.2 and 2.8 detectable sources for the 17- and 100-pulsar cases respectively.
In contrast, the current PPTA sensitivity curve produced by \citet{zhw+14} corresponds to $\lesssim10^{-4}$ 
detectable sources.  `Noise' caused by the summed GW 
signal from the binary SMBH population will further increase the difficulty of detecting individual binaries \citep[e.g.,][]{svv09,rwh+12}.

\subsection{GW memory bursts from coalescing binary SMBHs} 

A PTA data set with 20 pulsars timed with a precision of 100\,ns for 10\,yr is sensitive to memory 
bursts with amplitudes $h_{\rm mem}>5\times10^{-15}$ over $70-80\%$ of the data span \citep{vl10,cj12}. As the sensitivity of 
such an idealised PTA to memory bursts scales roughly as the square root of the number of pulsars \citep{vl10}, 
a PTA with 100 pulsars timed with 100\,ns precision for 10\,yr may be sensitive to memory bursts with $h_{\rm mem}>2\times10^{-15}$. 
However, our model suggests that only $\sim$$10^{-5}$ bursts with $h_{\rm mem}>5\times10^{-15}$ and $\sim$$10^{-3}$ bursts with 
$h_{\rm mem}>2\times10^{-15}$ are expected over 10\,yr. Thus, under the model presented here, GW memory bursts from coalescing binary 
SMBHs do not represent viable sources for PTAs.

\section{Discussion}

Our predictions for the GWB amplitude, $A_{\rm yr}$, are conservative within their respective scenarios, 
for a number of reasons. 
(\textit{i}) We do not account for minor galaxy mergers with stellar mass ratios $\mu_{*}<1/3$, 
or for mergers where the more massive galaxy has a mass $M_{*}<10^{10}M_{\odot}$. 
(\textit{ii}) We do not consider the possibility of gas accretion onto SMBHs prior to coalescence during galaxy mergers \citep[e.g.,][]{vvm+12}, which 
would raise the SMBH masses and hence the emitted GW amplitudes \citep[e.g.,][]{svc08}. 
(\textit{iii}) The most massive galaxies are typically 
found in cluster environments, where times between galaxy mergers may be shorter \citep[cf.][]{lpf+13}, implying a higher merger rate for these galaxies and hence 
a higher GW signal. However, we do not expect the inclusion of these factors to significantly affect our predicted GWB amplitudes.
We reiterate that the effects of interactions between binary SMBHs and their environments are unlikely to affect the predictions for the GWB amplitude  
at frequencies $f\gtrsim10^{-8}$\,Hz \citep{s13b,rws+14}, such as at $f_{\rm yr}$. This is because the orbital evolution of binary SMBHs radiating GWs at 
these frequencies is expected to be predominantly GW-driven, which further leads to the circularisation of the orbits.

Of all sources of uncertainty we consider in predicting the GWB amplitude given relevant observational quantities, the choice of 
galaxy merger timescale dominates the range of possible GWB amplitudes. Furthermore, the 
merger timescale may be even more uncertain than the range spanned by the predictions we consider \citep[L08;][]{kw08}. 
The simulations of L08 were conducted only for mergers between gas-rich disk galaxies, some of which contained small bulges, 
whereas we find that the GWB is likely dominated by binary SMBHs formed in mergers solely between early-type galaxies. 
Further theoretical studies of galaxy merger 
timescales for early-type systems are clearly required in order to better predict the GWB amplitude.
The dominance of low-redshift ($z\lesssim 1.5$) early-type major galaxy mergers of massive ($M_{*}\gtrsim5\times10^{10}M_{\odot}$) galaxies 
in determining the GWB amplitude is a further important consequence of our work for both theoretical and 
observational studies of galaxy mergers aimed at informing PTA research. 

The other significant source of uncertainty in our predictions is in the $M_{\bullet}-M_{\rm bul}$ relation, both in its local form and in its possible redshift-evolution. 
In contrast to uncertainty in the galaxy merger timescale, 
it is likely that this uncertainty will only be resolved through further observations which significantly expand the sample of known 
SMBH masses. Promisingly, \citet{dbc+13} report that hundreds of SMBH mass measurements may be possible with the Atacama Large Millimetre Array (ALMA).

Under what circumstances could the GWB amplitude lie outside the range we predict given all uncertainties that we consider? The predicted 
range of GWB 
amplitudes, $5.1\times10^{-16}<A_{\rm yr}<2.4\times10^{-15}$, encompasses all purely observational uncertainties, as well as uncertainty ranges that 
we set for other quantities for which observational constraints are poor, such as the galaxy merger timescale. 
It may be possible that these latter ranges are incorrect. 
Furthermore, not all galaxies may host a central SMBH, as we have assumed. The interaction between a binary SMBH and a third 
SMBH would likely cause the least massive SMBH to be ejected \citep[e.g.,][]{gs14}, lowering the number of coalescing SMBHs. If not every 
massive galaxy at $z\sim1$ formed with a central SMBH, the GWB amplitude would again be lowered. It may also be possible that binary SMBHs 
do not always coalesce on timescales less than the times between galaxy mergers.

The presence of a few strong GW emitters among the binary SMBH population implies that some excess, non-Gaussian scatter will be present in the 
GW signals produced by this population. The magnitude of this excess scatter in $A_{\rm yr}$ depends on exactly how many 
binary SMBH systems contribute significantly to the GWB. 
Using a semi-analytic galaxy formation model implemented in the Millennium simulation \citep{gwb+11}, \citet{rwh+12} suggested that the statistics of 
ToA variations induced by GWs from binary SMBHs are mildly non-Gaussian for frequencies $f>f_{\rm yr}/5$ because of appreciable 
contributions to the squared characteristic strain spectrum, $h_{c}^{2}(f)$, from individual binaries at every GW frequency.
Figure~5 shows the 
number of binary SMBHs in our fiducial model corresponding to galaxy mergers with primary stellar masses greater than or equal to a given 
$M_{*}$ (top), as well as the fractions of $A_{\rm yr}^{2}$ contributed by these binaries (bottom). 
We show in particular binaries radiating at a GW frequency of $f_{\rm yr}$ in a frequency bin of width $\Delta f=(10\,{\rm yr})^{-1}$.
Our fiducial model suggests that the contributions of 
individual GW sources to $h_{c}^{2}(f)$ are lower than estimated by \citet{rwh+12}. For example, the modelling in \citet{rwh+12} found that one source contributed 
$\sim50\%$ of $h_{c}^{2}(f)$ at a frequency of $2f_{\rm yr}/3$ in a frequency bin of width $(5\,{\rm yr})^{-1}$ (their Figure~2). In contrast, 
our empirical modelling in this paper suggests that the strongest $\sim400$ sources in such a frequency bin contribute $\sim50\%$ of $h_{c}^{2}(2f_{\rm yr}/3)$. 


\begin{figure}
\centering
\includegraphics[angle=-90,scale=0.58]{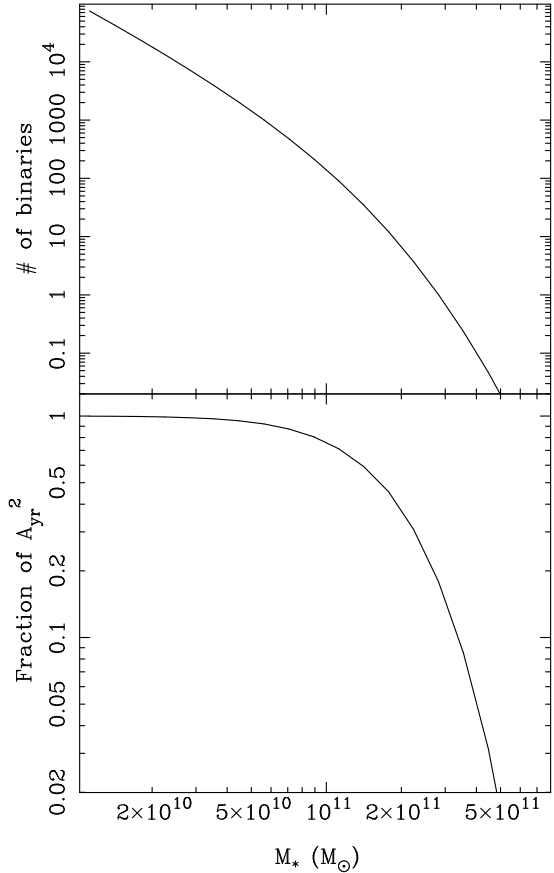}
\caption{\textit{Top:} The numbers of binary SMBH sources predicted by our fiducial model radiating at a GW frequency 
of $f_{\rm yr}$ in a frequency bin of width $\Delta f=(10\,{\rm yr})^{-1}$ at and above given values of $M_{*}$. \textit{Bottom:} 
The fractions of  $A_{\rm yr}^{2}$ contributed 
by binary SMBHs at and above given values of $M_{*}$. }
\end{figure}

This work and \citet{rwh+12} clearly predict different numbers of the most massive binary SMBHs. While this discrepancy will only be resolved with 
GW observations, we point out that the Schechter functions for the GSMFs 
that we use under-predict observed galaxy counts at the highest masses and the lowest redshifts \citep{bdl+12,mms+13}. 
Hence, it is possible that our model under-represents the contributions of the most massive binary SMBHs to the total GW signal. 
Differing typical galaxy merger mass ratios in cluster and field environments \citep[e.g.,][]{lpf+13} are a further complicating factor.

\section{Conclusions}

In this paper, we predicted the strength of the GWB from binary SMBHs and the occurrence of individual binary SMBH GW sources. 
Our approach was to use a selection of recent  
observational estimates for the average times between major mergers for galaxies with $M_{*}>10^{10}M_{\odot}$ and $z<3$ 
and for the GSMFs of early- and late-type galaxies in this mass and redshift range. We combined these quantities with 
empirical relations between galaxy and bulge stellar masses and between bulge and SMBH masses. 

We find that while current PTAs are unlikely to be sensitive to individual binary SMBHs, a PTA consisting of $\sim$100 pulsars timed with $\sim$100\,ns 
precision for 5\,yr will be sensitive to up to $\sim3$ binary SMBHs. Such a PTA may be achievable with the SKA \citep{l13}, but is 
possibly beyond the capabilities of FAST \citep{hdm+14}. Even such a PTA will, however, have a less than 0.1\% chance of detecting a 
GW memory burst from a coalescing binary SMBH. Thus, we conclude that while individual binary SMBHs may be detectable with a PTA based on 
next-generation radio telescopes, memory bursts from coalescing SMBHs are not likely to be detectable with any envisaged PTA. 

We predict that the characteristic strain amplitude of the GWB lies in the range   
$5.1\times10^{-16}<A_{\rm yr}<2.4\times10^{-15}$ with 95\% confidence, accounting for a variety of uncertainties. 
The upper end of the predicted amplitude range is equivalent to the 
best published 95\% confidence upper limit on the GWB amplitude \citep{src+13}. This reinforces the 
conclusion of \citet{src+13} that some models for the binary SMBH population that are consistent with current electromagnetic observations are 
already inconsistent with PTA constraints on the GWB. 
 
The dominant uncertainty in predicting the GWB amplitude appears to be caused by differences in theoretical predictions for the major merger timescale of massive 
galaxies. Higher values within our predicted range for $A_{\rm yr}$ correspond to the more commonly preferred choice of galaxy merger timescale 
(L08); GWB amplitudes $A_{\rm yr}>10^{-15}$ are within the sensitivity ranges of current and future PTAs. 
We strongly urge further work on quantifying the galaxy merger timescale, in particular for the mergers between massive early-type galaxies at redshifts 
$z<1.5$ which are likely to host the dominant contributors to the GWB. The other significant uncertainty in our predictions is in the local form and 
possible redshift-evolution of the $M_{\bullet}-M_{\rm bul}$ relation. PTA upper limits on or detections of the GWB may be able to meaningfully 
improve our knowledge of such otherwise poorly constrained facets of the formation and evolution of galaxies and SMBHs.

\section*{Acknowledgements}

The authors thank Alberto Sesana for useful discussions and Justin Ellis for sharing results on predicted sensitivity curves. 
The authors also acknowledge the comments of the anonymous referee, which helped to significantly improve the manuscript. 
V.R. is a recipient of a John Stocker Postgraduate Scholarship from the Science and Industry Endowment Fund and J.S.B.W. acknowledges an 
Australian Research Council Laureate Fellowship. GH is supported
by an Australian Research Council Future Fellowship. This work was performed on the swinSTAR supercomputer at the Swinburne University of Technology.

\end{document}